\documentclass[journal]{IEEEtran}

\usepackage{xcolor,soul,framed} 

\colorlet{shadecolor}{yellow}
\usepackage[pdftex]{graphicx}
\graphicspath{{../pdf/}{../jpeg/}}
\DeclareGraphicsExtensions{.pdf,.jpeg,.png}

\usepackage[cmex10]{amsmath}
\usepackage{array}
\usepackage{mdwmath}
\usepackage{mdwtab}
\usepackage{eqparbox}
\usepackage{url}
\usepackage{tabularx}
\usepackage{csquotes}
\usepackage{amsmath,amssymb}
\usepackage[colorlinks=true,bookmarks=false,citecolor=blue,urlcolor=blue]{hyperref} 
\usepackage[free-standing-units=true]{siunitx}
\usepackage{cite}
\begin{document}
\bstctlcite{IEEEexample:BSTcontrol}
    \title{Wavefront shaping and imaging through a multimode hollow-core fiber}
  \author{Zhouping~Lyu,~\IEEEmembership{Advanced Research Center for Nanolithography}\\
  ~\IEEEmembership{VU Amsterdam, Amsterdam, the Netherlands}\\
      Lyubov~Amitonova,~\IEEEmembership{Advanced Research Center for Nanolithography}\\
      ~\IEEEmembership{VU Amsterdam, Amsterdam, the Netherlands}
}


\maketitle

\begin{abstract}
Multimode fibers recently emerged as compact minimally-invasive probes for high-resolution deep-tissue imaging. However, the commonly used silica fibers have a relatively low numerical aperture (NA) limiting the spatial resolution of a probe. On top of that, light propagation within the solid core generates auto-fluorescence and Raman background, which interferes with imaging. Here we propose to use a hollow-core fiber to solve these problems. We experimentally demonstrate spatial wavefront shaping at the multimode hollow-core fiber output with tunable high-NA. We demonstrate raster-scan and speckle-based compressive imaging through a multimode hollow-core fiber.
\end{abstract}


\section{Introduction}
\IEEEPARstart{M}{ultimode} fibers (MMFs) enable high-resolution \emph{in vivo} imaging due to their inherent flexibility, simplicity, and minimal sample damage \cite{cao2023controlling,vasquez2018subcellular,turtaev2018high,ohayon2018minimally,wen2023single}. The ultimate goal of MMF-based imaging is to provide high-resolution high-contrast imaging through a compact sensor with a minimized footprint. Two common imaging methods through an MMF are raster-scan (RS) imaging, based on wavefront shaping (WFS) and optimized illumination, and compressive imaging (CI), based on random speckle illumination and computational reconstruction \cite{vellekoop2007focusing, vcivzmar2012exploiting, leite2021observing, katz2009compressive,pascucci2019compressive, amitonova2020endo,lochocki2023swept}. A conventional MMF usually consists of a silica core and a doped silica cladding. The pump laser produces Raman scattering while propagating through a fiber core \cite{wardle1999raman,stolen1980nonlinearity}. The laser beam also interacts with residual photoactive compounds of an optical fiber, leading to auto-fluorescence signal \cite{bianco2021comparative}. Therefore, Raman and auto-fluorescence signal of fused silica accumulated during light propagation through a fiber probe provides a strong background and decreases measurement sensitivity. Scattering and reflection at the interfaces of an MMF further increase the unwanted signal. Spectral filtering can help to partially reduce the background, but it also attenuates the fluorescent signal from the sample. The low signal-to-background ratio decreases the imaging contract and may totally destroy the fiber-based imaging of weak fluorescent samples.

Recently, hollow-core fibers (HCFs) have attracted a lot of interest \cite{komanec2020hollow,fokoua2023loss}. A hollow air-filled core offers unique properties for a wide range of applications: from gas laser and pulse compression to sensing, communication, and imaging \cite{balciunas2015strong, hassan2016cavity, yang2021hollow, wang2019hollow, lombardini2018high}. The small overlap between the guided light and the solid cladding material provides low auto-fluorescence and Raman background signal in comparison with standard fused silica waveguides \cite{luan2024situ,yu2016negative,yerolatsitis2019ultra}. The dispersion of the HCF is relatively low because of the absence of material dispersion \cite{fokoua2023loss}. HCFs can substantially improve the signal collection efficiency in nonlinear optical fiber endoscopy and have been shown to enable high-sensitive detection of the Raman signal~\cite{yerolatsitis2019ultra,klioutchnikov2020three,doronina2012raman}.

A hollow waveguide with internal metallic coatings is an attractive candidate among different categories of HCF \cite{jeliinkova2004hollow}. Unlike other HCFs, such as hollow-core photonic bandgap fibers and antiresonant HCFs, which are designed to guide a limited number of low-order modes resulting in a numerical aperture (NA) of less than $0.05$, an internal metallic hollow waveguide allows a continuum of modes to propagate, each with different attenuation coefficient \cite{szwaj2024double,komanec2020hollow}. In geometric ray optics, light propagating through an internal metallic hollow waveguide reflects off the dielectric film. Lower-order modes reflect fewer times, while higher-order modes reflect more, leading to faster attenuation of higher-order modes. Therefore, the internal metallic HCF can offer a high-NA that depends on the fiber length and diameter. Although HCFs offer many advantages over traditional solid-core fibers, the investigation of the NA of HCFs as multimode hollow-core fibers (MHCFs) and computational imaging through MHCFs has not been addressed.

Here we experimentally demonstrate the wavefront shaping through an MHCF. We generate and investigate diffraction-limited foci with an NA up to $0.42$ at the MHCF output. Furthermore, we experimentally demonstrate high-resolution imaging through the MHCF by two approaches: RS imaging with optimized illumination and CI with random speckle illumination.

\section{Methods}
\subsection{Experimental setup}

\begin{figure}[t]
\centering
\includegraphics[width = 0.5\textwidth]{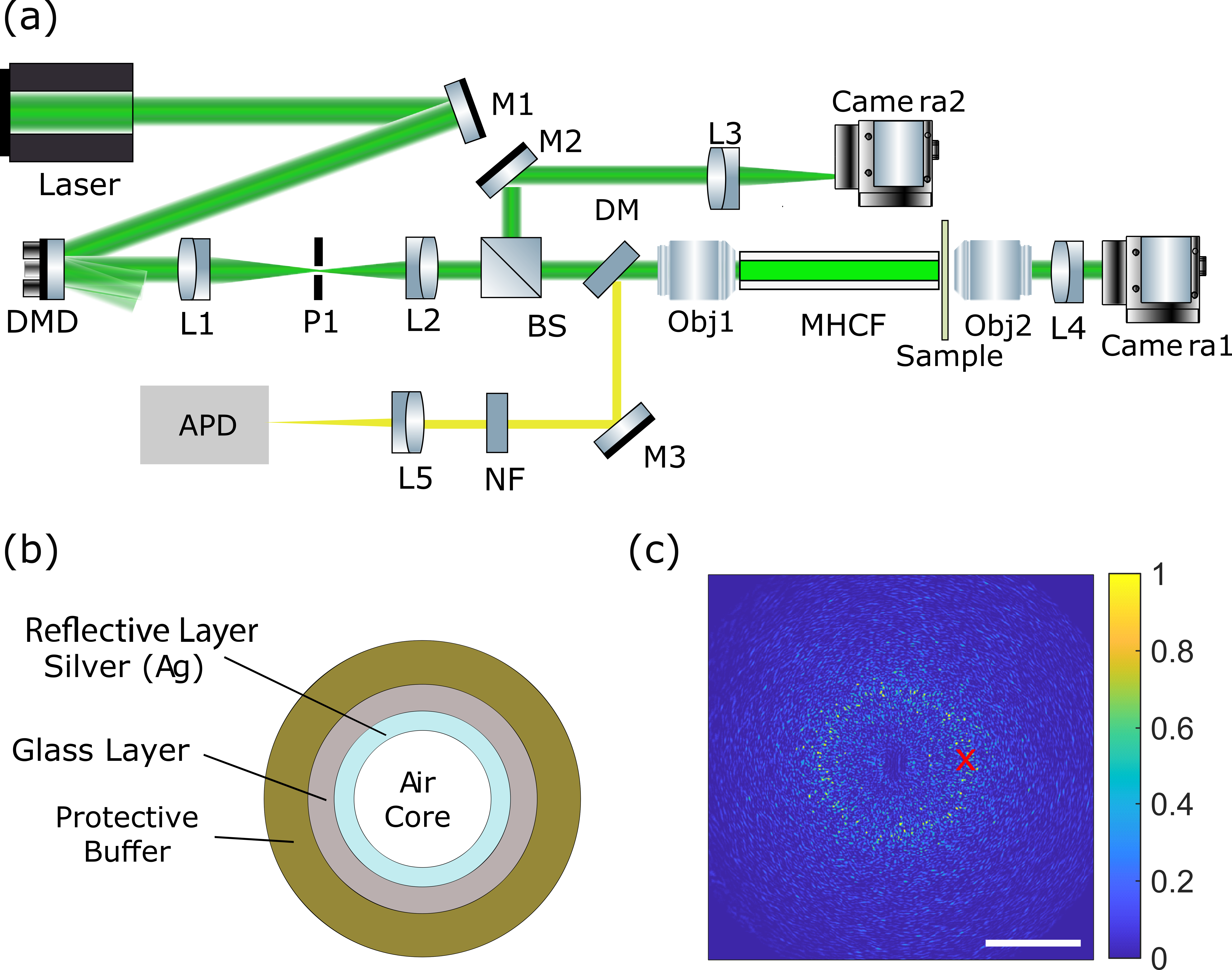}
\caption{\label{Experiment setup}{(a) Schematic of the experimental setup. DMD, Digital Micromirror Device; L, lens; P, pinhole; M, mirror; BS, beam splitter; DM, dichroic mirror; Obj, Objective; NF, notch filter; APD, avalanche photodiode detector. (b) The cross-section of the multimode hollow-core fiber (MHCF). (c) Speckle pattern of MHCF output if the input light is focused at the location marked by the red cross. Core diameter is 500~$\mu$m. The scale bar is 100~$\mu$m.}}
\end{figure}

Figure~\ref{Experiment setup}(a) shows the experimental setup. We use the bare MHCFs that consist of a hollow silica tube with a highly reflective silver (Ag) coating deposited on the inner surface with the length of \SI{7}{cm} and core diameters of $500$ and $300~\mu$m (Guiding Photonics). The cross-section of the MHCF probe is shown in Fig.~\ref{Experiment setup}(b). The mode profiles of the internal metallic MHCF are similar to that of a step-index MMF and can be determined by solving Maxwell's equations with the appropriate cylindrical boundary conditions \cite{Marcatili1964HollowMA}. However, the propagation constants of modes in MHCFs are complex numbers, indicating that all modes are leaky. In theory, the attenuation constant $\alpha = 0.5 n_0 k_0 d^2 \left ( 1-R \right )U$, where $U$ is the transverse phase constant, $R$ is the reflection coefficient, $d$ is fiber diameter, $k_0$ is the wavevector, and $n_0 = 1$ is the refractive index of the core \cite{jeliinkova2004hollow, miyagi1984wave}. High-order modes, with larger $U$ and smaller 
$R$, have higher transmission losses. Thus the NA and the loss depend on the fiber length, fiber diameter, and the NA of the input coupling objective.

We use a continuous-wave light source with a wavelength of \SI{532}{\nm} [Cobolt Samba] as a pump and expand it to a beam diameter of $8$ mm. After being reflected by a mirror (M1), the laser beam is incident to the Digital Micromirror Device (DMD) from Texas Instrument driven by the DLP V-9501 VIS module (Vialux). The $-1$st diffraction order of the DMD passes through the pinhole (P1), and the phase modulation of the DMD is imaged at the pupil plane of the input objective (Obj1) by the 4f system consisting of two lenses, L1 (f1 = $150$~mm) and L2 (f2 = $100$~mm). Throughout the experiments, we used three different objectives to test the NA of the MHCF:
\begin{itemize}
    \item \text{Olympus RMS10X}: 10$\times$, NA = 0.25;
    \item \text{Olympus RMS20X}: 20$\times$, NA = 0.4;
    \item \text{Olympus RMS20X-PF}: 20$\times$, NA = 0.5.
\end{itemize}
The output of the MHCF is imaged on Camera1 by an output objective (Obj2, Leica, 63x, NA = 0.7) and a tube lens (L4, f4 = 250 mm). A sample is put at the output plane of the HCF. The fluorescent signal from the sample is collected by the same MHCF, separated from the pump beam by a dichroic mirror (DM), and recorded by an avalanche photodiode detector (APD, Thorlabs) with a lens (L5, f5 = \SI{60}{mm}). A notch filter (NF) is used to further remove the input light from the fluorescence signal.

\subsection{Spatial wavefront shaping}
Similar to a step-index MMF, coherent light coupled into an MHCF is decomposed into fiber modes. The output can be calculated as the sum of all these propagated, weighted, leaky modes, which results in a speckle pattern. Figure~\ref{Experiment setup}(c) shows the output of the MHCF with a core diameter of $500~\mu$m. The light is coupled to the MHCF at the input facet position marked by the red cross, using an \text{Olympus RMS10X} objective with an NA of $0.25$. The bright ring structure with higher intensity is observed at the same radial position as the input light. A DMD has been used for spatial WFS. A DMD chip contains thousands of micromirrors arranged in a grid, each capable of rotating $\pm 12^{\circ}$ to switch between \enquote*{on} and \enquote*{off} states. Arranging the \enquote*{on} and \enquote*{off} state of the micromirrors as a grating. Lee hologram encodes the relative phase shift to the spatial shift between the adjacent gratings~\cite{lee1979binary}. Compared with the liquid crystal SLMs, DMDs give a higher beam shaping fidelity with a faster modulation rate~\cite{turtaev2017comparison}. The optimized wavefronts are calculated by a feedback-based algorithm and then projected onto the DMD, creating the targeted light distributions on the fiber output~\cite{vellekoop2007focusing}. 

We use the dual reference algorithm to create a focal spot at the MHCF output facet\cite{mastiani2021noise}. 
In our experiment, the Hadamard order for each group is $10$, the number of overlap segments is $60$, and $3$ phases are used for each Hadamard pattern, resulting in $6144$ phase patterns projected to the DMD and $6144$ images acquired by the camera. The camera has a limited frame rate of up to $100$ Hz. Therefore, pre-calibration requires around $1$ minute for data collection. The optimized wavefront for each focus position on the fiber output was individually calculated based on the Fourier transform of the feedback intensities and the Hadamard transform of the segments. In the next step, these optimized wavefronts are projected to the DMD one by one to get the foci at the output of the MHCF.

\subsection{Imaging through a hollow core fiber}
We experimentally implement two modalities of computational imaging through an MHCF: WFS-based RS imaging and random speckle-based CI. RS imaging uses WFS to generate foci at any desired position at the fiber output. A sequential scan of these foci is performed in a zigzag pattern within a square frame. For each focal spot illumination, the fluorescent signal from the sample is collected by the same MHCF, filtered out by the DM, and then detected by the APD.  

In contrast to RS imaging, which uses an optimized wavefront at the input facet of the MHCF to create a focal spot at the sample plane, CI employs a 2D scanning at the input facet of the fiber and uses random speckles to illuminate the sample. For the calibration step, we perform a 2D scan of a $M = 30 \times 30$ grid within an area of about $35 \times 35$~${\mu m}^{2}$ at the input facet of the MHCF. Each input point, after propagation through the HCF, generates a speckle pattern, which illuminates the sample. The $M$ speckle patterns are recorded as images of $N \times N$ pixels in the calibration step, using the Camera1. The illumination patterns are transformed into $M$ row vectors with a size of $1 \times N^2$. By stacking these row vectors, we create an $M \times N^2$ measurement matrix $\boldsymbol{A}$.
In the measurement step, the same 2D scan is performed on the input of MHCF. The total fluorescence response of a sample for each speckle pattern is collected by the same fiber and measured with an avalanche photodiode. The total intensities of $M$ speckle patterns are measured, building a $M \times 1$ signal vector $\boldsymbol{y}$.

The CI process can be described using a simple equation $\boldsymbol{A}\boldsymbol{x} = \boldsymbol{y}$, where $\boldsymbol{x}$ is the flattered $N^2 \times 1$ sample vector, which is unknown. The compressive sensing algorithms are used to solve the underdetermined equation while $M \ll N^2$. The problem is solved assuming the sample sparsity on a specific basis. We choose total variation \texttt{TV} minimization \texttt{tval3} algorithm \cite{li2013tval3}, which uses the gradient sparse prior information with equality constraints. The isotropic denoising model is chosen: $\underset{\boldsymbol{x}}{\text{min}}\sum_{i}\left \| D_{i}\boldsymbol{x} \right \|,$ such that~$\boldsymbol{A}\boldsymbol{x} = \boldsymbol{y}, \boldsymbol{x} >=0$, where $D_{i}\boldsymbol{x}$ is the discrete gradient of $\boldsymbol{x}$ at pixel $i$ and $\left \| \cdot \right \|$ is $l_2$-norm. The \texttt{tval3} is chosen because the \texttt{TV} algorithms gradient sparse assumption gives a sharp edge and smooth field and the \texttt{tval3} has a relatively fast speed.

\section{Results}

\begin{figure}[t]
\centering

\includegraphics[width = 0.5\textwidth]{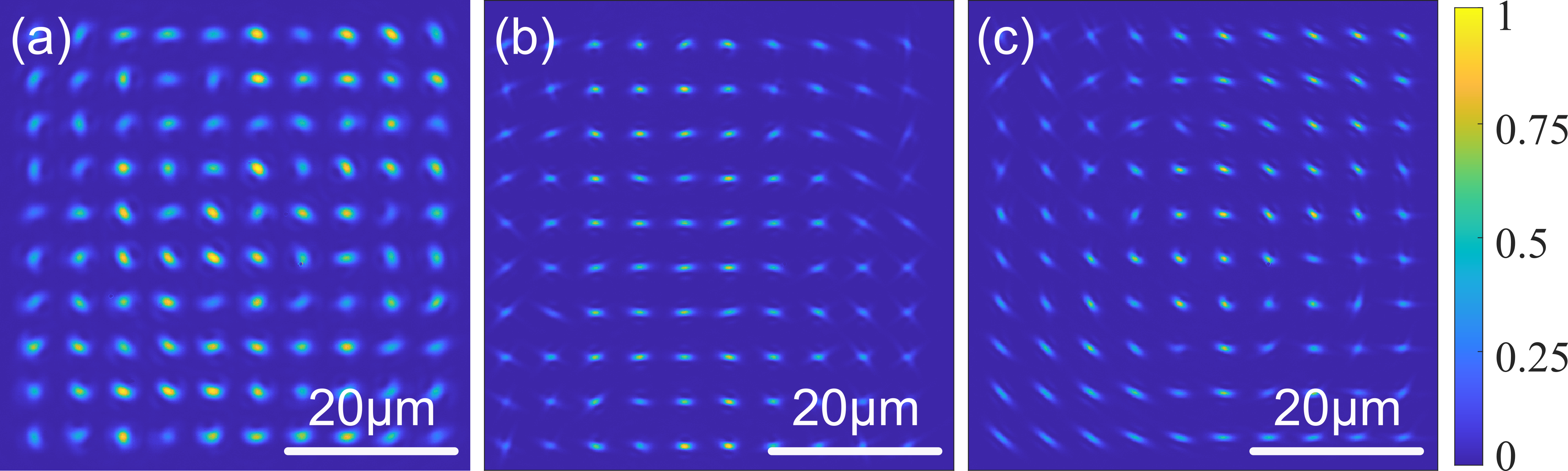}
\caption{\label{wfs}{Wavefront shaping through a hollow fiber with a core diameter of $500~\mu$m. The incoherent sum of independently generated focal spots arranged in a $10 \times 10$ grid around the center of an MHCF. The presented field of view is $55.8~\mu$m by $55.8~\mu$m. Objective on the fiber input has NA of $0.25$~(a), $0.4$~(b), and $0.5$~(c). The scale bars are 20~$\mu$m.}}
\end{figure}

In the first set of experiments, we demonstrate the WFS through the MHCF. We use a hollow fiber with a core diameter of 500~$\mu$m and the coupling microscope objective \text{Olympus RMS10X} with NA of $0.25$.
Multiple foci at different locations arranged in a $10 \times 10$ grid have been optimized around the center of the MHCF using a single WFS procedure described above. The distance between two optimized focal spots on the MHCF output is $5.3~\mu$m. The incoherent sum of the foci within a field of view of $55.8~\mu$m by $55.8~\mu$m is presented in Fig.~\ref{wfs}(a). The WFS allows us to precisely control the light on the output of hollow fiber.

In the second set of experiments, we compare the performance of WFS through the same MHCF for different coupling NAs. We investigate the NA range from $0.25$ to $0.5$. We repeat the experiments for three input objectives as described in the experimental setup section to experimentally change the input NA. The results are presented in Fig.~\ref{wfs}. The focal spots have been optimized at the same positions on the output fiber facet.
The incoherent sum of all the foci for the input objectives with NA of $0.25$, Fig.~\ref{wfs}(a), $0.4$, Fig.~\ref{wfs}(b), and $0.5$, Fig.~\ref{wfs}(c), are shown. We can see that the higher input NA leads to a smaller size of focal spots after the WFS.

We characterize the NA of the generated foci for different input objectives by analyzing their full width at half maximum (FWHM). The FWHM of the short axis, $W_s$, for different focal spots is extracted by independently fitting each focal spot by a 2D tilted Gaussian function.
The NA is then calculated assuming diffraction-limited foci as NA$_f = \lambda/(2 W_s)$, where $\lambda$ is the wavelength of the input light.
The averaged NA of $100$ foci on the MHCF output generated by WFS for 3 different input objectives are presented in the second column of Table~\ref{NA}.
We also calculate the NA of speckles within the same field of view of $55.8~\mu$m by $55.8~\mu$m. We couple light to the MHCF using different objectives and different input beam positions and record speckle patterns on the fiber output. The averaged two-dimensional (2D) spatial power spectrum for each objective is calculated by averaging the 2D fast Fourier transform of 900 speckle patterns generated with the given objective. The cutoff frequency of the averaged spatial power spectrum, $\nu_\text{cutoff}$, is identified at the point where the slope of the curve's descent noticeably decreases.
The NA is then calculated by $\nu_\text{cutoff} = 2 \text{NA}_f/\lambda$.
The estimated NAs of speckle patterns for different input NAs are presented in the third column of Table~\ref{NA}. The calculated NA for the hollow fiber output derived from speckle patterns and optimized foci match very well, confirming the accuracy of the measurements. 
For the first two input objectives with NA of $0.25$ and $0.4$, the calculated NA of the hollow fiber output is close to the NA of the objective. However, for the input objective with an NA of $0.5$, the estimated NA is noticeably lower. This can be explained by more rapid attenuation of higher-order modes within a hollow fiber.

\begin{table}[b]
\centering
\begin{tabularx}{0.49\textwidth}{X X X}
\hline
 Input NA  & Foci & Speckles\\ \hline
 0.25  & 0.22 & 0.23\\
 0.4  & 0.39 & 0.37\\
 0.5  & 0.41 & 0.42\\ \hline
\end{tabularx}
\caption{\label{NA}The experimentally estimated NAs of WFS-based foci and speckles for three objectives with different input NAs.}
\end{table}

\begin{figure}[t]
\centering
\includegraphics[width = 0.5\textwidth]{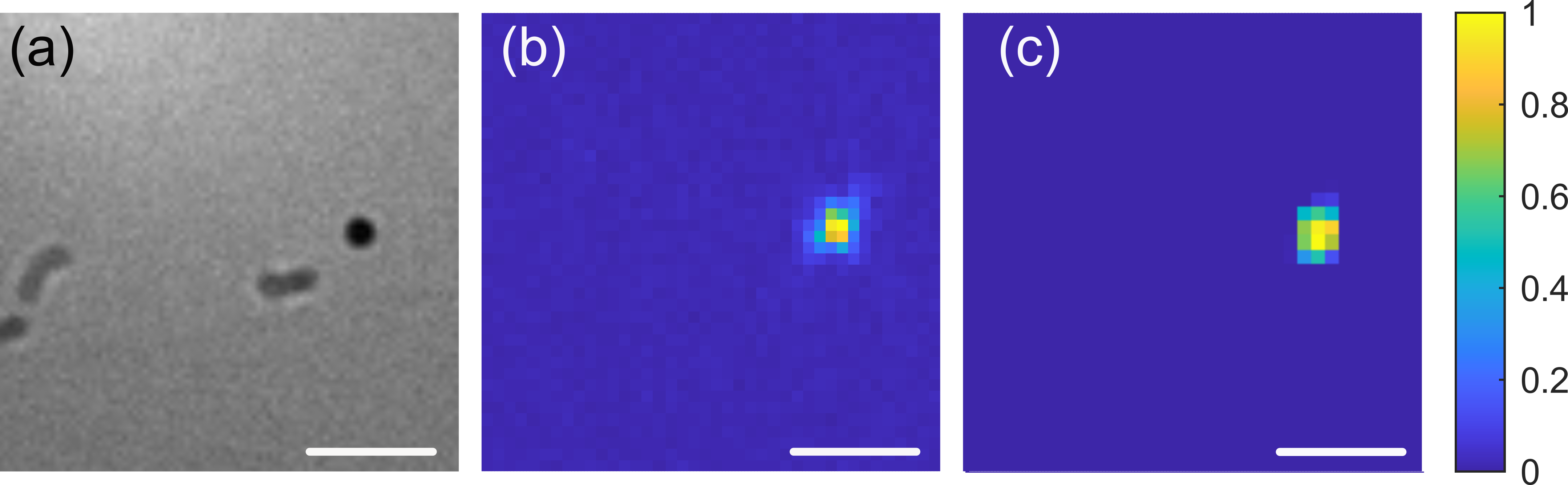}
\caption{\label{imaging}{(a) Bright-field reference image of $1.14~\mu$m fluorescent bead. Images of the same sample obtained through a hollow fiber probe by RS imaging (b) and CI approach~(c). The scale bars are 2~$\mu$m.}}
\end{figure}

In the final set of experiments, we demonstrate the potential of an MHCF to serve as a high-NA imaging probe. We experimentally demonstrate RS imaging and CI through a hollow fiber. 
We use an MHCF with a diameter of 300~$\mu$m and the input objective with an NA of $0.4$ and a pupil plane diameter of 7.2 mm (\text{Olympus RMS20X}). As a sample, we use a fluorescent microparticle (PS-FluoRed-1.0) with a diameter of $1.14~\mu$m. As a reference, the bright-field image of the sample is captured using an objective with NA of 0.7 and Camera1. The result is presented in Fig. \ref{imaging}(a).
For the RS imaging, a $40 \times 40$ foci arranged in a grid with a distance of $0.175~\mu$m is optimized using WFS as described above. Each focal spot is sequentially projected on the fiber output and used to illuminate the sample at a certain location. As a result, we can scan illumination across the sample. The result of RS imaging is presented in Fig.~\ref{imaging}(b). The measurement speed of RS imaging is limited by the DMD fresh rate, which is up to $23$~kHz and can reach $14$ frames per second with $40 \times 40$ focal spots per frame.

For the CI approach, a set of $900$ speckle patterns illuminates the sample individually. During the reconstruction, the speckles are cropped to a field of view at the center of fiber of $7.2~\mu m \times 7.2~\mu m$ and resized to $33 \times 33$ pixels. The result of CI is shown in Fig. \ref{imaging}(c). With the same DMD fresh rate, CI can achieve $25$ frames per second.

\vspace{0.5cm}

\section{Conclusion}
We experimentally demonstrated spatial WFS through an MHCF. We investigated the NA of MHCF-based imaging. 
We demonstrate that the NA on the output of an MHCF can be more than $0.4$ for an MHCF with a diameter of $500~\mu$m and a length of $7$ mm. We successfully performed raster-scan imaging and compressive imaging through an MHCF. Our results show the potential of MHCF being a high-NA low background imaging probe.

\vspace{1cm}
\bibliographystyle{IEEEtran}
\bibliography{IEEEabrv,Bibliography}

\vfill


\end{document}